\documentclass{optica-article}

\journal{opticajournal} 

\articletype{Research Article}

\usepackage{ulem}
\usepackage{upgreek}

\begin{document}

\title{Aspheric Lens Design Proposal for near-perfect mode-matching of a Broadband Quantum Dot Micropillar to a Single-Mode Fibre}

\author{Yichen Zhang,\authormark{1,3} David Dlaka,\authormark{1,*} James McDougall,\authormark{1,2} James Y Tai,\authormark{1} Petros Androvitsaneas,\authormark{1} Edmund Harbord,\authormark{1} Ruth Oulton,\authormark{1} and Andrew B. Young\authormark{1}}

\address{\authormark{1} Quantum Engineering Technology Labs, H. H. Wills Physics Laboratory and School of Electrical, Electronic, and Mechanical Engineering, University of Bristol, BS8 1FD, UK\\
\authormark{2} Quantum Engineering Centre for Doctoral Training, H. H. Wills Physics Laboratory and School of Electrical, Electronic, and Mechanical Engineering, University of Bristol, BS8 1FD, UK\\
\authormark{3} Currently with Department of Electrical Engineering, University of Cambridge, Cambridge, CB3 0FA, UK}

\email{\authormark{*}david.dlaka@bristol.ac.uk} 


\begin{abstract*} 
Quantum dots in micropillars are one of the most promising options for a bright, deterministic single photon source. While highly efficient devices (>95\%) have been designed, there remains a significant bottleneck that impacts the overall system efficiency: the large numerical aperture of the output mode. This leads to inefficient coupling of emitted photons into single-mode fibre, thus limiting practical integration into quantum computing and communication architectures. We show that with the addition of a well designed aspheric $\mathrm{SiO_2}$ microlens we can decrease the mode-matching losses to a SMF from 83.1\% to <0.1(0.1)\%. This can result in a single photon source design with 96.4(0.1)\% end-to-end efficiency, paving the way for scalable photonic quantum technologies.

\end{abstract*}

\section{Introduction and Methodology} 

The development of an on-demand single photon source (SPS) with near-ideal efficiency and indistinguishability could accelerate a number of photonics-based quantum technologies \cite{ObrienQTech2009,CouteauQTech2023}, such as the ability to reliably generate large linear cluster states \cite{KLM2001} for applications in quantum computing \cite{KokQComp2007}, as all-optical memory-less repeaters \cite{EconomouRepeaters2017}, as a source in secure quantum communications and cryptography \cite{BB84,JenneweinRNG2000,BeveratosSinglePhotonQCrypt2002}, generating high-N00N states, as well as other applications in quantum metrology \cite{DowlingNoon2008,AfekNoon2010,CouteauQSens2023}. Among the various sources of quantum light, epitaxial quantum dots in micropillars \cite{MoreauPillars2001,PeltonPillarOG2002,VuckovicPillars2003} have been a particularly strong contender due to the high demonstrated efficiencies at the first lens (internal efficiency $\xi$) \cite{Somaschi2016,DingPillars2016,UnsleberPillars2016,GinesDlaka2022}, and deterministic light-matter interactions \cite{PetrosPRB2016,PetrosACS19}. However, due to the fact that the emission from the micropillars is over a large solid angle, efficiently mode-matching to a single mode fibre (SMF) requires additional engineering. Some examples include embedding the quantum dot in a Solid Immersion Lens (SIL) structure \cite{SartisonSIL2017,ChenSIL2018,AhnDotSIL2023}, or mode-matching with various hemispherical lens or circular Bragg grating (CBG) structures on either the sample or on the collection fibre \cite{LochnerSIL2019,BremerLenses2022,Schwab2022CouplingLE}. These approaches are able to shape the output mode and reduce losses due to mode mismatch to <20-10\%, but negatively affect the internal efficiency $\xi$. In order to push the end-to-end efficiency further, open cavity designs have been demonstrated \cite{WarburtonLensedRecord2021}, with the current state of the art achieving 71\% into SMF with 97.95\% single-photon purity \cite{JianweiLensed2023}.

In this work, we address the issue of near-lossless fibre coupling of ultra-efficient SPSs by taking previously optimised micropillar cavity designs with internal efficiency $\xi=96\%$ \cite{DlakaPillars2024} and integrating an aspheric SiO$_2$ microlens on top; by controlling the lens profile, the mode field diameter and numerical aperture (NA) of the pillar emission can be shaped. Using finite-difference time-domain (FDTD) simulations, we demonstrate a lens resulting in a near-unity ($>99\%$) overlap between the emitted mode of the micropillar and an industry standard single-mode fibre, whilst maintaining the internal efficiency of our original micropillar design.

\begin{figure}[ht!]
\centering\includegraphics[width=0.85\linewidth]{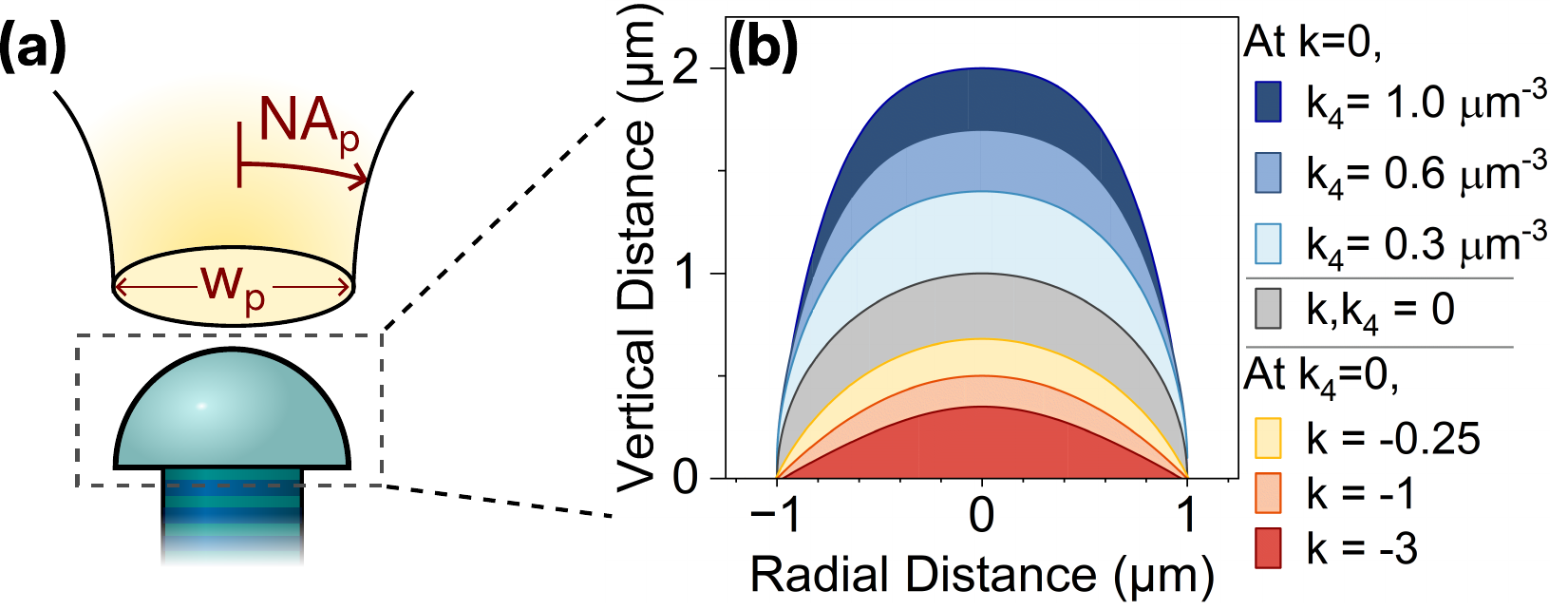}
\caption{(a) The emission of a micropillar with a microlens on the top face. The mode field diameter $w_\mathrm{p}$ and numerical aperture $\mathrm{NA_p}$ are the key parameters in determining the coupling efficiency of the emission into SMF; (b) The cross-section of various aspheric lenses shows the hemispherical $\mathrm{k}=\mathrm{k_4}=0$ case in grey become increasingly hyperbolic ($\to$ red) as $\mathrm{k}\to-\infty$ or increasingly quadric ($\to$ blue) as $\mathrm{k_4}\to\infty$.}
\label{LensIntroFig}
\end{figure}

We use a commercial FDTD solver to simulate telecom O-band GaAs/AlAs micropillars under the same simulation conditions stated in \cite{DlakaPillars2024}, where the glass material used for the additional lenses in this work is $n_{\mathrm{SiO}_2}=1.45$. The emission from the top of the micropillars undergoes a fast Fourier transform which allows for resolution of emitted modes in momentum space. The emission profile of the micropillar-lens systems, which is illustrated in Figure \ref{LensIntroFig}(a), is described by the pillar Mode Field Diameter (MFD), labelled $w_\mathrm{p}$, and the Numerical Aperture (NA) of the emission as defined by the $1/e^2$ criterion, labelled $\mathrm{NA_p}$. These values, as determined from a 2D planar monitor placed above the pillar or lens, are compared to the well-defined Gaussian of an industry standard single-mode fibre (SMF-28) with MFD $w_\mathrm{f}\approx9.2\, \upmu$m and $\mathrm{NA_f}=0.14$ at $\lambda\approx1.3\, \upmu$m. To minimise the efficiency losses due to mode-matching, we initially use the near field MFD $w_\mathrm{MFD}$ to inform whether any telescopic magnification is required, which would affect the optimal NA of the emission. This interdependence of $w$ and $\mathrm{NA}$ follows
\begin{equation}
w_\mathrm{f}\approx\frac{2\lambda}{\pi \mathrm{NA_f}} \label{magEq}
\end{equation}
for small NA (see equation 4.3 in \cite{YanBook2019}). In this work we match the MFDs of the micropillar-lens device and a SMF, thereby avoiding the need for telescopic magnification. Therefore, following the method in \cite{ KataokaCoupling2010}, the comparison of the NA in the far-field (momentum) from the device and a SMF can be used to calculate a coupling (mode-matching) efficiency $\eta_\mathrm{SMF}$ as
\begin{equation}
    \eta_\mathrm{SMF} = \frac{4 \mathrm{NA^2_p} \mathrm{NA^2_f}}{\left(\mathrm{NA^2_p}+\mathrm{NA^2_f}\right)^2} \label{SMFEfficiencyEq}
\end{equation}
where $\mathrm{NA}_f$, $\mathrm{NA}_p$ are the far-field numerical apertures of the fibre and pillar beams (per the $1/e^2$ criterion).

A glass microlens with an aspheric profile can be used to change the pillar MFD and NA. The aspheric lens profile follows the equation
\begin{equation}
    z(x) = \frac{x^2/\mathrm{R}}{1+\sqrt{1-(1+\mathrm{k})\frac{\mathrm{x}^2}{\mathrm{R^2}}}} + \mathrm{k_4}\,x^4 \label{lensEq}
\end{equation}
where $\mathrm{R}$ is the lens radius at the base, $\mathrm{k}$ is the conic coefficient, and $\mathrm{k_4}$ is the quadric coefficient; higher order coefficients have been omitted to keep the optimisation space to only these three dimensions. Some example profiles for $\mathrm{R}=1\ \upmu$m have been shown in Figure \ref{LensIntroFig}(b). The $\mathrm{k,k_4}=0$ case results in a hemisphere (grey) which becomes more hyperbolic (red) with the conic coefficient $\mathrm{k}$, while the $\mathrm{k_4}x^4$ contribution leads to more bullet-shaped (blue) lens profiles. N.B. The axes are shown in terms of the lens height and width, but the coordinate system origin per Eq.\ref{lensEq} is centred at the top of each lens.

\section{Design Proof Of Concept}

The demand for additional mode matching tools, such as the aspheric lens technique proposed in this work, becomes evident when considering the emission profile from simple micropillars. A simple design strategy \cite{DlakaPillars2024}, where the micropillar diameter can be carefully selected to Purcell suppress leaky decay, can be used to find families of high-efficiency high-bandwidth devices at small diameters ($<2-3\, \upmu$m). However, these devices have a small mode field diameter (MFD) $\mathrm{w_p}\sim 1.5-2.5\,\upmu$m and a large angular divergence with NA $\sim0.5-0.8$. To mode-match this emission to the profile of a standard O band SMF-28, which has a MFD $\mathrm{w_f}\approx9.2\, \upmu$m and NA$_\mathrm{P}\approx0.14$, is extremely challenging and involves magnification using multiple, possibly custom, bulk optical lenses. These will introduce excess losses and the setup would need to be changed for each micropillar design. Alternatively, an additional integrated mode-matching element on each device can shape the output mode such that the emission from each pillar has the MFD and NA intrinsically matched to a SMF. This trivialises the mode-matching problem, removing the need for magnification, and opening up the possibility of a direct-to-fibre butt coupling. Here, we show that this transformation can be done by modifying the profile on an integrated aspheric lens, providing a scalable solution to fine-tune the emission for ultra-efficient collection in a SMF. 

\begin{figure}[ht!]
\centering\includegraphics[width=0.65\linewidth]{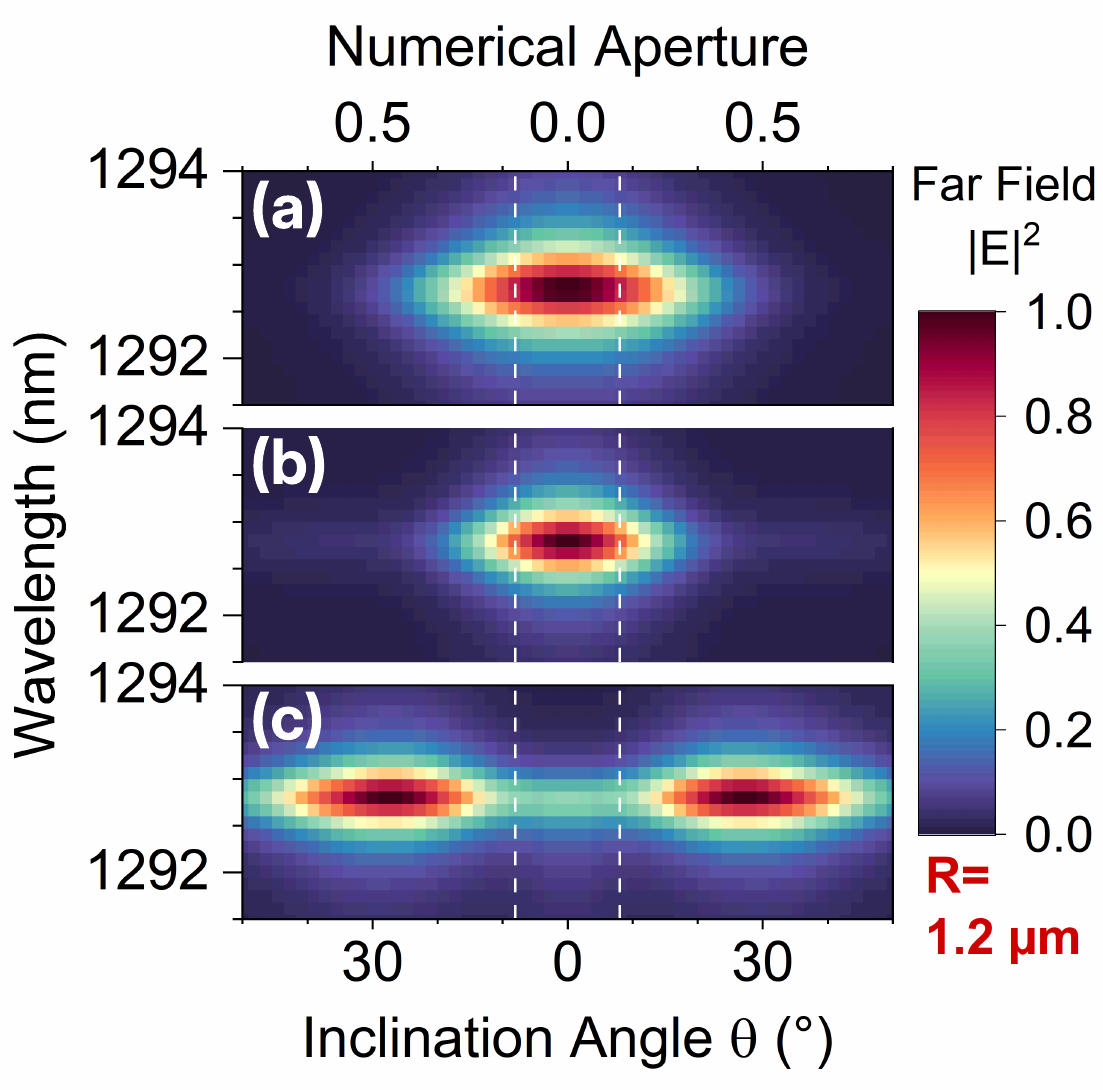}
\caption{The far-field projection of the emitted modes is shown as a function of the wavelength and inclination angle from the emission normal for a lens with radius $\mathrm{R}=1.2\, \upmu$m. The aspheric lenses have quadric parameter $\mathrm{k_4}=$ (a) 0.25, (b) 0.75, and (c) 1.5 $\upmu\mathrm{m}^{-3}$, representing an under-, maximally-, and over-corrected device, respectively. The white dashed lines mark the SMF $\mathrm{NA_f}=0.14$.}
\label{changingK4}
\end{figure}

We begin the optimisation of the aspheric lens design for a $\xi=96\%$ efficient micropillar as described in \cite{DlakaPillars2024}. This pillar design has 13 (35.5) top (bottom) distributed Bragg reflector (DBR) GaAs/AlAs layers and a side-leakage suppressing diameter $d\approx1.9\ \upmu$m $(\equiv5\lambda/n)$. The lens is a SiO$_2$ aspheric lens, centred on and contacting the top surface. Using the three free parameters of the aspheric lens equation (\ref{lensEq}) - the conic coefficient $\mathrm{k}$, the quadric coefficient $\mathrm{k_4}$, and the lens radius $\mathrm{R}$ - we attempt to match the pillar MFD and NA, $w_\mathrm{p}$ and $\mathrm{NA_p}$, to that of a SMF, $w_\mathrm{f}$ and $\mathrm{NA_f}$. Given the large disparity between $\mathrm{NA_p}$ and $\mathrm{NA_f}$ our design strategy is to find a lens that minimises $\mathrm{NA_p}$. The first parameter we investigate is $\mathrm{k_4}$, which contributes to the height of the lens and gives it a smooth truncated cone profile. Figure \ref{changingK4} highlights the effect on the far-field emission, where the $|\mathrm{E}|^2$ is shown as a heat-map in momentum space plotted against wavelength (y-axis) and inclination angle (x-axis), As $\mathrm{k_4}$ increases from 0.25 $\upmu\mathrm{m}^{-3}$ (a) to 0.75 $\upmu\mathrm{m}^{-3}$ (b), the emission NA decreases to more closely match the fibre $\mathrm{NA_f}=0.14$ which appears as a white dashed line. After reaching a minimum in the NA in (b), further increasing $\mathrm{k_4}$ leads to a non-Gaussian bimodal angular distribution, as shown for $\mathrm{k_4}=1.5\,\upmu\mathrm{m}^{-3}$ in Fig.\ref{changingK4}(c). This demonstrates that by modifying $\mathrm{k_4}$ we can find a local minimum for the NA.

Performing a complete investigation of the $\mathrm{k_4}$ values that minimise the NA at various lens radii $\mathrm{R}$ and $\mathrm{k}$, we find the pillar $\mathrm{NA_p}$ is minimised at $\mathrm{k}=0$ for all $\mathrm{R}$ and $\mathrm{k_4}$ (See Supplementary Material). This simplifies the design challenge allowing us to set $\mathrm{k}=0$ and focus on two free parameters: the lens radius $\mathrm{R}$ and the quadric parameter $\mathrm{k_4}$.

\begin{figure}[ht!]
\centering\includegraphics[width=1\linewidth]{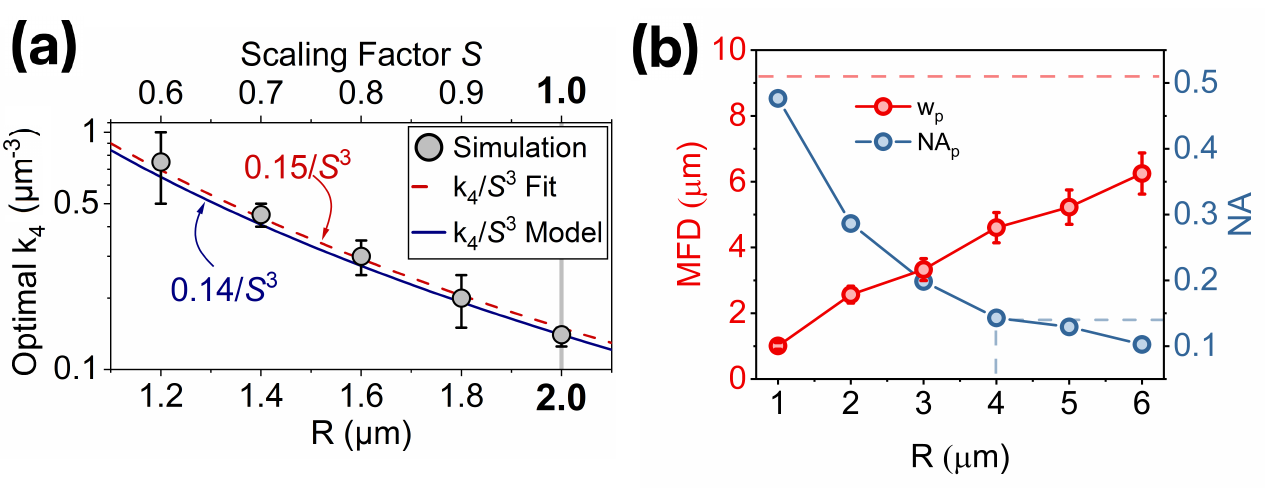}
\caption{A plot of the $\mathrm{k_4}$ values which minimise the emission NA as a function of radius. The FDTD simulated $\mathrm{k_4}$ values are shown in grey. Starting from the $\mathrm{R}$=2$\,\upmu$m, $\mathrm{k_4}=0.14\,\upmu $m$^{-3}$ lens, the projection of $\mathrm{k_4}$ values using the magnified lens relationship stated in (\ref{k4ModelEq}) is shown in blue. A fit of the data (red) has a fitted $\mathrm{k_4}=(0.149\pm0.003)\, \upmu\mathrm{m}^{-3}$ demonstrating excellent agreement. (b) The mode field diameter, left axis, and numerical aperture, right axis, for lenses with minimised NA at different radii $\mathrm{R}$. The dashed red and blue lines correspond to the MFD and NA of a SMF, respectively.}
\label{scalingMfdNa}
\end{figure}

Given we have found the minimum NA for one lens radius $\mathrm{R}$, we explore the effect of changing the radius on the NA. We find that lenses with different radii that minimise the NA all have the same lens profile, but scaled. To demonstrate this, we consider a lens scaled by a factor $S$ which will have a new radius $\mathrm{R'=S\,R}$. Applying this transformation to the system, it follows that the scaled lens has a corresponding $\mathrm{k_4'}$ which satisfies
\begin{equation}
\mathrm{k'_4}(\mathrm{R}')=\frac{\mathrm{k}_4}{\left(\mathrm{R}'/\mathrm{R}\right)^3} = \frac{\mathrm{k_4}}{\mathrm{S}^3}
\label{k4ModelEq}
\end{equation}
Figure \ref{scalingMfdNa} shows a plot of the FDTD-determined $\mathrm{k_4}$ values resulting in the smallest NA as a function of the lens radius R as grey data points. The blue solid line shows the model in Eq. \ref{k4ModelEq} where the $\mathrm{R}=2\, \upmu$m, $\mathrm{k_4}=0.14\upmu\mathrm{m}^{-3}$ has been taken as an arbitrary $\mathrm{S}=1$ starting point. The excellent agreement of the model with the simulated data, and with the fitted value of Eq.\ref{k4ModelEq} shown as a dashed red line with $\mathrm{k_4}=0.149(0.003)\upmu\mathrm{m}^{-3}$, suggests that the profile of aspheric lenses which minimise the emission NA is indeed invariant under a scaling transformation.

Using the scaling equation we are able to explore the impact of the lens radius on the micropillar emission. Figure \ref{scalingMfdNa}(b) shows the MFD $\mathrm{w_p}$ and $\mathrm{NA_p}$ of the micropillar emission as a function of lens radius $\mathrm{R}$. The SMF MFD $\mathrm{w_f}$ appears as a horizontal red dashed line, while the fibre NA $\mathrm{NA_f}$ as a blue dashed one. The blue vertical dashed line corresponds to the lens radius where perfect mode-matching is achieved in the far-field. Note that in the near-field, the mode-matching does not occur within the radius range simulated. We find that our design strategy of minimising $\mathrm{NA_p}$ is sub-optimal, as when we match the NA ($\mathrm{NA_f}=\mathrm{NA_p}$) the MFD is too small ($\mathrm{w_p}<\mathrm{w_f}$). For optimised SMF coupling we need to slightly modify our design.

\section{Optimal Aspheric Lens Design for $<0.1\%$ SMF Coupling Loss}

\begin{figure}[htbp!]
\centering\includegraphics[width=0.7\linewidth]{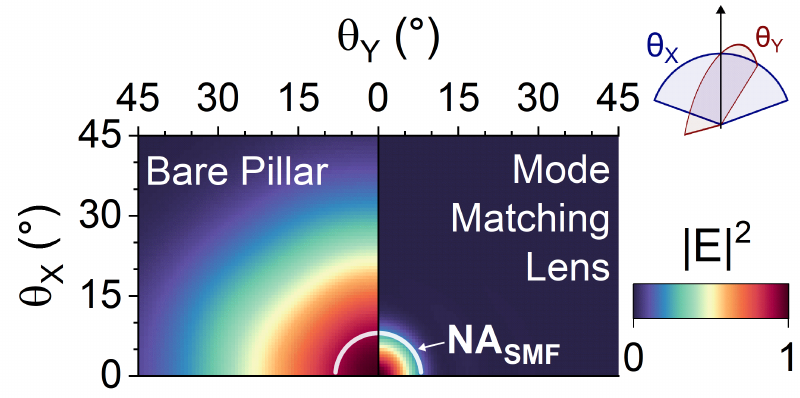}
\caption{The far-field profiles of the field intensity $|\mathrm{E}|^2$ are shown as a function of angle from the normal along the x and y axes ($\theta_x, \theta_y$ as shown in the inset) for the lensless bare micropillar and the optimised $\mathrm{R}=5.7\ \upmu$m, $\mathrm{k}=0$, $\mathrm{k_4}=3.75\times10^{-3}\, \upmu\mathrm{m}^{-3}$ lens. The NA of a SMF is shown as a white circle.}
\label{BestLensFig}
\end{figure}

We have already established that modifying the $\mathrm{k_4}$ allows us to tune the NA, and the lens radius allows us to tune the MFD. However, the two are not completely decoupled and we find some interdependence. By reducing the $\mathrm{k_4}$ we can increase the NA, but also slightly increase the MFD. We are now able to apply a small perturbation to the lens profile in Fig.\ref{scalingMfdNa} which allow us to simultaneously mode-match the MFD and NA. We find the optimal lens has $\mathrm{k}=0$, $\mathrm{R}=5.7\ \upmu$m, and $\mathrm{k_4}=3.75\times10^{-3}\, \upmu\mathrm{m}^{-3}$. The far field emission profile is shown in Fig. \ref{BestLensFig}, where the significant reduction in the spread of the emitted mode can be seen in the side-by-side comparison with a bare pillar. The mode after modification closely matches the NA of the SMF, which is illustrated with a white line. Due to the fact that $\mathrm{w_p}=\mathrm{w_f}$, one can use (\ref{SMFEfficiencyEq}) to find $\eta_\mathrm{SMF}=0.999(0.1)$; in other words, that the fibre coupling losses due to mode mismatch are $\approx0.1(0.1)\%$, or $4.35\times10^{-3}$ dB. It's important to note that this near-perfect mode-matching to an industry standard SMF is achieved without incurring any losses to the internal efficiency $\xi$

\begin{figure}[htbp!]
\centering\includegraphics[width=0.5\linewidth]{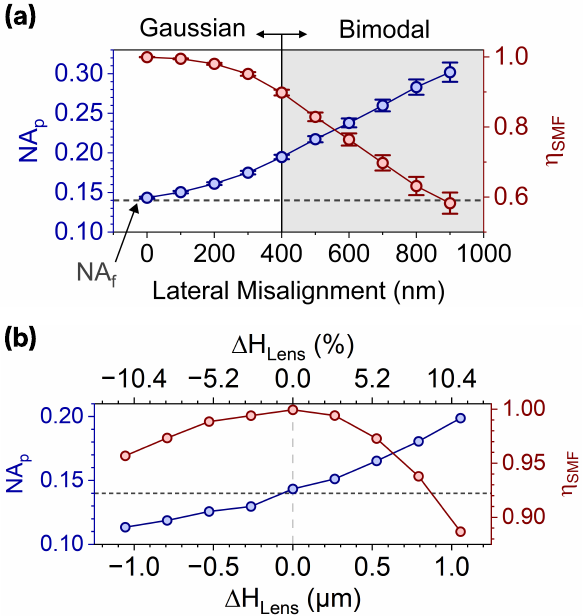}
\caption{(a) The device NA (left y-axis, blue) and mode-matching efficiency (right, red) to a SMF as a function of lateral lens offset for the $\mathrm{R}=5.7\ \upmu$m, $\mathrm{k}=0$, $\mathrm{k_4}=3.75\times10^{-3}\, \upmu\mathrm{m}^{-3}$; (b) The device NA (left, blue) and SMF mode-matching efficiency $\eta_\mathrm{SMF}$ (right, red) are shown as a function of fabrication tolerance quantified as an error in the lens height $\Delta\mathrm{H}$ in units of $\upmu$m on the bottom x-axis, and as a percentage on the top axis. The grey dashed lines mark the SMF NA.}
\label{ToleranceFig}
\end{figure}

Finally, it is important to demonstrate that these designs are robust to fabrication errors. Considering the micropillar has a small diameter $\sim1.9\, \upmu$m, the alignment of the lens to the pillar is of particular interest. The tolerance on this is examined in Figure \ref{ToleranceFig}(a), where we show the pillar emission $\mathrm{NA_p}$ (blue, left y-axis) and corresponding SMF coupling efficiency $\eta_\mathrm{SMF}$ (red, right y-axis) as a function of lateral misalignment between the pillar and lens centres. At 0 nm offset, the $\mathrm{NA_p}\simeq\mathrm{NA_f}$, which is marked by a grey dashed line; as a lateral offset is introduced (<400 nm), the emission NA is broadened such that $\eta_\mathrm{SMF}$ >90\%. When the offset is severe (>400 nm), the mode becomes non-Gaussian (bimodal) and the mode-matching breaks down. The 300-400 nm tolerance on the lateral misalignment is generous and well within tolerance for fabrication techniques, such as two-photon polymerisation (TPP) or focused ion beam (FIB) etching \cite{ReitzensteinLensPrint2017, schiappelli2004efficient}.

We also examine the tolerance to errors in the lens shape/profile, simulated as deviations in $\mathrm{k_4}$ from the optimal value; we parametrise this as an error in the lens height $\Delta\mathrm{H_{Lens}}$. In Figure \ref{ToleranceFig}(b), we show plots the pillar emission $\mathrm{NA_p}$ (blue, left) and $\eta_\mathrm{SMF}$ (red, right) as functions of $\Delta\mathrm{H_{Lens}}$ which is expressed in $\upmu$m on the bottom x-axis and as a \% on the top x-axis. It can be seen that within a significant $\pm10\%$ error on the lens height, corresponding to $\approx\pm1\, \upmu$m, the SMF coupling efficiency $\eta_\mathrm{SMF}$ remains above 90\%. This is a demonstration of strong resilience to fabrication tolerances on the quadric term of the aspheric. The results in Figure \ref{ToleranceFig} show that aspheric lenses on micropillars are robust to fabrication errors and should be manufacturable using standard clean-room processing.

\section{Conclusion}

In this work we have highlighted the need to take a holistic approach to micropillar design that also considers the deployment of the source into a larger network of quantum devices. For this, an efficient cavity design is not sufficient; it is paramount that the output modes are compatible with industry standard components, such as telecom fibres. We have proposed a simple strategy for matching the Gaussian mode emitted from the top of $>96\%$ efficiency O-band micropillars to that of a single-mode fibre. By integrating a glass aspheric lens on top of the GaAs/AlAs pillar to modify the near- and far-field emission, the coupling efficiency of the micropillar emission into single-mode fibre, $\eta_\mathrm{SMF}$, can be increased from $16.9\%$ to $99.9(0.1)\%$, with ultra-small coupling losses on the order of $\sim10^{-3}$ dB. 

With this proposed aspheric lens designs, we can achieve ultra-efficient SMF coupling with no impact to the internal efficiency of the micropillar. This drastically simplifies the bulk optics required for mode-matching, which are responsible for most of the end-to-end losses, and opens a route to direct-to-fibre coupling in the near field. Experimental realisation would constitute a significant technological advancement which could open the door for more advanced applications of solid state single photon sources in quantum computing, cryptography, communication, and metrology.

\begin{backmatter}
\bmsection{Funding}
QuantERA (EP/Z000491/1); Engineering and Physical Sciences Research Council (META-TAMM: EP/X029360/1, QECDT: EP/S023607/1).

\bmsection{Acknowledgment}
The authors would like to thank James Mallord and Junyang Yan for their contributions to this work.

\bmsection{Disclosures}
The authors declare no conflicts of interest.

\bmsection{Data Availability Statement}
Data underlying the results presented in this paper are not publicly available at this time but may be obtained from the authors upon reasonable request.
\end{backmatter}\\

\noindent See Supplement 1 for supporting content.
\bibliography{ASPHERIC_LENS_ON_PILLAR}

\end{document}


\maketitle

 The optimisation of the aspheric lens design begins with the 13/35.5 micropillar at the fifth $d_5$ optimal diameter ($d\approx1.9\ \upmu$m) from \cite{DlakaPillars2024} as a $\xi=96\%$ efficient pillar, where a SiO$_2$ aspheric lens is centred on and contacting the top surface. We use the proportion of the mode inside NA=0.14, labelled $\eta_{0.14}$, as a criterion which allows for quantitative comparison between Gaussian and non-Gaussian profiles, such that
\begin{equation}
    \eta_{0.14} = \frac{\int_0^{0.14}|\mathrm{E}(\theta)|^2 \ \mathrm{d}\theta} {\int_0^{1}|\mathrm{E}(\theta)|^2\  \mathrm{d}\theta} \label{eta14eq}
\end{equation}
where the integral limits are expressed as NA=$\sin(\theta)$ for convenience. 
 
\begin{figure}[ht!]
\centering\includegraphics[width=1\linewidth]{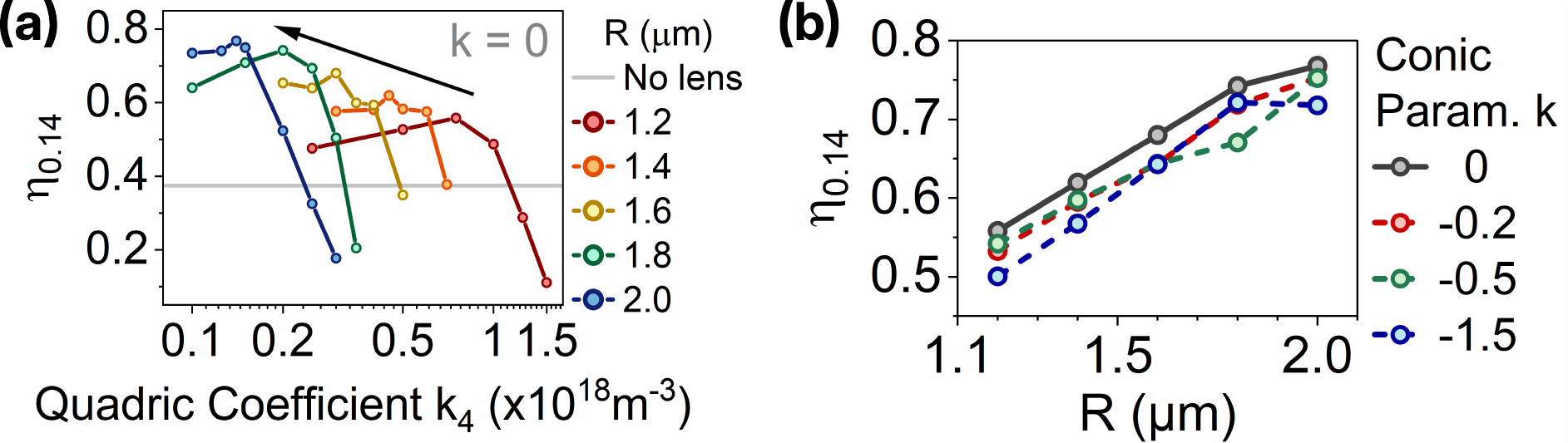}
\caption{(a) The $\eta_{0.14}$ values are shown as a function of the quadric $\mathrm{k_4}$ coefficient for lenses with $\mathrm{k=0}$ and varying $\mathrm{R}$, showing a presence of a local maximum with a dependence on the lens radius. The grey line represents the no-lens (bare pillar) case and the solid arrow shows the increasing peak $\eta_{0.14}$ value for larger radii; (b) The $\eta_{0.14}$ values for aspheric lenses with varying $\mathrm{R}$ and conic coefficient $\mathrm{k}$ are shown, where for each data point the $\eta_{0.14}$ maximising $\mathrm{k_4}$ value was determined and used, showing that the $\mathrm{k}=0$ maximises performance.}
\label{ApFig1}
\end{figure}

 We start from the simple $\mathrm{k}=0$ case with a non-zero $\mathrm{k_4}$ component, recording the changes in $\eta_{0.14}$ as the $\mathrm{k_4}$ contribution is increased. This data is shown for lenses with $\mathrm{R}=1.2-2\, \upmu$m in Figure \ref{ApFig1}(a). Observations of the trend for a particular radius (for instance $\mathrm{R}=1.2\ \upmu$m shown in red) reveal three notable features: firstly, an initial increase of the field within NA=0.14 up to a local maximum ($\mathrm{k_4}\approx0.75\, \upmu$m$^{-3}$) before a sharp and sudden decline to below that of the bare pillar, marked by a dotted line where $\eta_{0.14}\approx0.35$; secondly there is a dependence of $\mathrm{k_4}$ on $\mathrm{R}$ where larger lenses require a lower $\mathrm{k_4}$ parameter. Thirdly, the local maxima of $\eta_{0.14}$ becomes higher at larger lens base radii $\mathrm{R}$. This indicates that the minimal NA obtainable by each lens is inversely proportional to it's radius, and the requirement of mode-matching to a certain fibre (in this context, an SMF-28 with NA = 0.14) place a lower bound on the lens radius.
 
Next, we consider optimisations over the conic coefficient $\mathrm{k}$. A plot of the $\eta_{0.14}$ values obtained for the $\mathrm{k}=0-1.5\, \upmu\mathrm{m}^{-3}   ;$ lenses in the $\mathrm{R}=1.2-2.0\ \upmu$m range is shown in Figure \ref{ApFig1}(b). The $\mathrm{k_4}$ value used for each individual data point is the value which maximises $\eta_{0.14}$ by sweeping $\mathrm{k_4}$ for each $\mathrm{k}$, at each radius $\mathrm{R}$. It is evident that hyperbolic deviations from the hemispherical base, caused by an increasingly negative conic coefficient k, are less effective in narrowing the far-field emission.

The key conclusions from the investigation into the mode-shaping capabilities of the integrated aspheric lens are that the best lens which can most effectively narrow the far field NA has a hemispherical base with $\mathrm{k=0}$, and that there is an inverse relationship between the lens radius and the smallest NA obtainable.

\bibliography{ALoP_SUPPLEMENTARYMATERIALS}